\documentstyle[sprocl,epsfig]{article}

\def\Journal#1#2#3#4{{#1} {\bf #2}, #3 (#4)}

\def\NPB{{\em Nucl. Phys.} B}
\def\PLB{{\em Phys. Lett.}  B}
\def\PRL{\em Phys. Rev. Lett.}
\def\PRD{{\em Phys. Rev.} D}

\def\EPJC{{\em Eur. Phys. J.} C}
\def\NPPS{{\em Nucl. Phys.} B (Proc. Suppl.)}
\def\PRT{\em Phys. Rept.}

\def\beqn{\begin{equation}}
\def\eeqn{\end{equation}}
\def\barr{\begin{eqnarray}}
\def\earr{\end{eqnarray}}
\def\etal {{\it et al}. }
\def\eg {{\it e.g}. }

\def\rvac {|0\rangle}
\def\lvac {\langle 0|} 

\begin{document}

\title{$N^*$ Spectrum in Lattice QCD
\footnote[2]{This work was done in collaboration with
T.\ Blum, P.\ Chen, N.\ Christ, M.\ Creutz, C.\ Dawson, G.\ 
Fleming, R.\ Mawhinney, S.\ Ohta, S.\ Sasaki, G.\ Siegert, A.\ Soni, 
P.\ Vranas, M.\ Wingate, L.\ Wu and Y.\ Zhestkov 
(RIKEN-BNL-Columbia-KEK collaboration).}}

\author{Shoichi Sasaki}

\address{RIKEN BNL Research Center,
Brookhaven National Laboratory,\\ Upton, NY 11973, USA\\ E-mail:
ssasaki@bnl.gov}   


\maketitle\abstracts{We investigate the mass of the parity partner $N^*
(1/2^{-})$ of the nucleon $N(1/2^{+})$, in lattice QCD
using a new lattice discretization scheme for fermions,  domain wall
fermions (DWF).  DWF possess exact chiral symmetry and flavor
symmetry, both of which are required for this problem, even at finite
lattice spacing.  Our calculation reproduces the large mass splitting
between those two states, in good agreement with experiment.  We
also present preliminary results for the mass of the positive-parity
excited state $N'(1/2^{+})$.}

\section{Introduction}

\indent\indent
It is important to understand the hadron mass spectrum from the first
principles of quantum chromodynamics (QCD), the fundamental theory of the
strong interactions. As is well known, the only method for such a
first-principle calculation is numerical lattice QCD.  Recent lattice
calculations of the light-hadron mass spectrum in the quenched
approximation agree well with experimental values 
within about 5\%~\cite{CP-PACS}.  However, this success is restricted 
to ground states and does not apply to excited states.  
The excited-state mass spectrum is
big challenge in lattice QCD.  

In this study, we focus on a negative-parity nucleon excited-state
$N^*(1535)$,  which is theoretically identified as the {\it parity
partner} of the nucleon $N(939)$.  We are especially interested in the 
mass splitting between $N$ and $N^*$ from the viewpoint 
of parity partners. As is well known, the mass splitting between parity 
partners would be absent if chiral symmetry were preserved~\cite{Pagels}.  
In other words spontaneous chiral-symmetry breaking is
responsible for the absence of such parity doubling in the actual 
hadron spectrum. In this sense, regardless of a model or a theory, 
it is important to handle the {\it chiral symmetry
and its spontaneous breaking} for reproducing 
precisely the mass splitting between parity-partner hadrons. 

Indeed, both non-relativistic quark models~\cite{Isgur}
and bag models~\cite{Bag}, which explicitly break chiral symmetry,
fail to reproduce the large mass splitting 
between $N(939)$ and $N^*(1535)$~\cite{PDG}.  
The non-relativistic quark models are based on a harmonic
oscillator description of the orbital motion of constituent quarks.  
The plausible value of its oscillator quantum should be a couple 
of hundred MeV to reproduce the observed 
charge radius and magnetic moment of the nucleon~\cite{Isgur}.
Since this model regards $N^*$ as a state with one quantum excitation 
in orbital motion, it indicates that $N^*$ lies just 
a few hundred MeV above the ground state.
Even worse, we face a serious problem of the wrong ordering 
of $N^*(1535)$ and the positive-parity excited state 
$N'(1440)$~\cite{PDG}
because $N'$ should be assigned two oscillator quanta in this 
model~\cite{Isgur}.
Furthermore this wrong ordering cannot be improved by the conventional 
one-gluon-exchange potential model of the residual interaction between 
constituent quarks~\cite{Glozman}. In addition, this difficulty led 
Glozman and Riska to propose another candidate for the residual 
interaction~\cite{Glozman}. It is easy to see that the bag models face 
essentially the same problem~\cite{Bag}.

It is clearly an interesting question whether lattice QCD can 
reproduce this large $N$-$N^*$ mass splitting.  
However, conventional lattice fermion schemes had some 
difficulty in this challenge. The Nielsen-Ninomiya 
no-go theorem~\cite{Nielsen} 
dictates that either chiral symmetry or flavor symmetry or both
are supposed to be violated at finite lattice spacing, 
while both of them are essential in this subject. 
The Wilson fermions explicitly break chiral symmetry at finite lattice
spacing and hence are quite inadequate for the current problem.  
Although Kogut-Susskind (KS) fermions have a remnant $U(1)$ chiral 
symmetry at finite lattice spacing, they are still not capable of the
$N^*$ mass calculation.  The main reason is that KS fermion has only 
discrete flavor symmetries which belong to a subgroup of the $SU(4)$ 
flavor symmetry~\cite{Golterman}.  
There are only three irreducible representations, {\bf8}, {\bf 8'} 
and {\bf 16} for KS baryon operators due to this incomplete
flavor symmetry. Two appropriate representations {\bf 8} and {\bf 16}, 
to which $N^*$(1535) belongs, involve also $\Lambda$(1405),
$\Lambda$(1520) and $N$(1520)~\cite{Golterman}.  
The study of $N^*$ spectrum using KS fermions always 
faces this inevitable contamination from lower mass states.

Several years ago, Kaplan~\cite{Kaplan} advocated a new 
type of lattice fermion scheme, and Shamir~\cite{Shamir,DWF} 
reformulated it for lattice QCD simulations.
They are called domain wall fermions (DWF), 
which utilize a fictitious fifth dimension.
An important feature of DWF is that chiral symmetry is almost 
exactly preserved even at the finite lattice spacing.  
This is achieved because the symmetry violation 
is suppressed exponentially in terms of the number of lattice sites 
$N_{s}$ in the extra dimension~\cite{DWF}. 
In other words, $N_s$ gives us a way to control the violation.  
Recent lattice calculations with DWF have shown that
good chiral properties are obtained for moderate $N_s$ like 16 
if the lattice spacing is small enough, like 0.2 fm~\cite{DWF}.  
In addition, the flavor symmetry is also well preserved in this 
fermion discretization. 
Thus we are led to an attempt to use DWF for lattice QCD calculations 
of the $N^*$ mass spectrum~\cite{Sasaki}.

\section{Baryon operators}

\indent\indent
The mass $m_{B}$ of the low-lying baryon $B$ is extracted from the 
two-point correlation function 
composed of the baryon interpolating operator ${\cal O}_{B}$:
%
%
\beqn
G_{{\cal O}_{B}}(t)=\sum_{\vec{x}}\lvac T\{ {\cal O}_{B}(\vec{x},t) 
{\bar{\cal O}}_{B}(0,0) \}\rvac \;,
\eeqn
which behaves like $\exp(-m_{B}t)$ for large $t$ 
since $G_{{\cal O}_{B}}(t)$ is projected out at zero 
spatial momentum through the sum over $\vec x$~\cite{Fucito}.
We focus on the nucleon channel, the spin-half iso-doublet baryons. 
There are two possible interpolating operators for the
corresponding quantum number $J^P = 1/2^+$ even if we restrict them
to contain no derivatives and to belong to the
$(\frac{1}{2},0)\oplus(0,\frac{1}{2})$ 
chiral multiplet under $SU(2)_{L}\otimes SU(2)_{R}$~\cite{Cohen}:
%
%
\barr
B^{+}_{1}&=&\varepsilon_{abc}(u^{T}_{a}C\gamma_{5}d_{b})u_{c}\;, \\
B^{+}_{2}&=&\varepsilon_{abc}(u^{T}_{a}Cd_{b})\gamma_{5}u_{c}\;,
\earr
where $abc$, $ud$, $C$ and $\gamma_{5}$ have usual meanings as color,
flavor, charge conjugation and Dirac indices. 
The superscript ``$+$'' denotes the positive parity.

The operator $B_{1}^+$ is preferred for use in lattice QCD
to extract the signal of the nucleon ground-state. 
On the other hand, the operator $B_{2}^+$ is conventionally discarded 
since $B_{2}^+$ is expected to couple weakly to the nucleon ground-state 
due to the vanishing non-relativistic limit~\cite{Leinweber}. 
In our calculation, the nucleon interpolating operator is assigned 
to $B_{1}^+$ in the conventional way. We try to calculate the 
excited-state mass spectrum from $B_{2}^+$. 

Multiplying the left hand side of the previous positive parity operators 
by $\gamma_{5}$, we obtain the interpolating 
operators with negative parity~\cite{FXLee}:
%
%
\barr
B^{-}_{1}&=&\gamma_{5}B^{+}_{1}=\varepsilon_{abc}(u^{T}_{a}C\gamma_{5}d_{b})
\gamma_{5}u_{c}\;, \\
B^{-}_{2}&=&\gamma_{5}B^{+}_{2}=\varepsilon_{abc}(u^{T}_{a}Cd_{b})u_{c}\;.
\earr
The point to notice is the relation between the nucleon 
two-point functions with opposite parities
%
%
\beqn
G_{B^{+}}(t) = -\gamma_{5}G_{B^{-}}(t)\gamma_{5}\;.
\label{eq:posneg}
\eeqn
This means that the two-point correlation function can couple 
to both positive and negative parity states~\cite{Fucito}:
%
%
\barr
G_{B^{+}}(t)&\rightarrow& \;\;\;(1+\gamma_{t})G_{0}(t) + (1-\gamma_{t})
G_{0}(-t)\;, \\
G_{B^{-}}(t)&\rightarrow& -(1-\gamma_{t})G_{0}(t) - (1+\gamma_{t})
G_{0}(-t)\;,
\earr
where $G_{0}(t)=\theta(t)A_{+}e^{-M_{+}t}+\theta(-t)A_{-}e^{+M_{-}t}$
\footnote[2]{In the quenched approximation,
$N^*$ can be regarded as the stable baryon like the nucleon.}.
$M_{\pm}$ denote masses of the opposite parity state.
Since $G_{0}(t)$ possesses the contribution from 
both positive and negative parity lowest-lying states~\cite{Fucito},
which propagate in the opposite time direction respectively,
we are threatened by contamination 
from the backward propagating opposite parity state in the case of 
the baryon spectrum with some boundary condition.
An appropriate boundary condition is required in 
the time direction to reduce the contamination of the opposite parity 
state.

At the end of this section, let us briefly review the knowledge that 
non-broken chiral symmetry imposes parity doubling in 
the hadron spectrum~\cite{{Pagels},{Cohen}}. 
For the sake of simplicity, we consider a particular 
transformation of the $SU(2)$ chiral symmetry as 
$[{\cal Q}_{5}, u]=+i\gamma_{5}u$ and $[{\cal Q}_{5}, d]=-i\gamma_{5}d$.
Then, one can easily find that the two-point correlator composed of each
$B_{1}$ and $B_{2}$ should transform as
%
%
\beqn
\left[ {\cal Q}_{5}, B^{+}_{1,2}(x){\bar 
B}^{+}_{1,2}(0)\right]= 
i \left\{ \gamma_{5}, B^{+}_{1,2}(x){\bar B}^{+}_{1,2}(0)\right\}
\label{eq:trans}
\eeqn
in the chiral limit; $\left[{\cal Q}_{5}, H\right]=0$. 
Suppose that the vacuum possesses chiral symmetry: ${\cal Q}_{5}\rvac = 0$. 
Eq.(\ref{eq:trans}) shows that
the two-point correlation function 
of the nucleon and $\gamma_{5}$ have to anti-commute,
%
%
\beqn
\left\{ \gamma_{5}, G_{B^{+}}(t)\right\} = 0\;.
\eeqn
Immediately, Eq.(\ref{eq:posneg}) gives
%
%
\beqn
G_{B^{+}}(t)=G_{B^{-}}(t) \;,
\eeqn
which means that parity doubling arises 
in the nucleon channel due to chiral symmetry~\cite{Cohen}. 
Of course, the chiral symmetry is spontaneously 
broken, ${\cal Q}_{5}\rvac \ne 0$ in the QCD vacuum so that such a 
parity doubling never occurs in the actual spectrum~\cite{Pagels}. 
In this sense, the spontaneous breaking of chiral symmetry is 
responsible for the absence of parity doubling.
\footnote[3]{This argument ignores the consequence of the 't Hooft 
anomaly condition~\cite{tHooft} and also spoils the possibility of 
the massless baryon because $G_{B^+}(t)$ and $G_{B^-}(t) $ are
defined at zero spatial momentum.}

\section{Numerical results}

\indent\indent
We generate quenched QCD configurations on a $16^3 \times 32$ 
lattice with the standard single-plaquette Wilson action 
at $\beta=6/{g^2}=6.0$. The quark propagator is calculated by 
using domain wall fermions with a fifth dimension of $N_{s}$=16 
sites and the domain-wall height $M$=1.8. 
According to our simulations~\cite{RBC},
this corresponds to lattice 
units $a^{-1}\approx 1.9$ GeV from $am_{\rho}$=0.400(8) in
the chiral limit and spatial lattice size $La \approx 1.7$ fm.

In this work, we provide forward-type and backward-type quark 
propagators with the wall source at two different locations ($t$ = 5 and 27)
and take the average of two measurements
for the hadron spectrum in each configuration.
We use 205 independent gauge configurations for the lightest 
two quark masses, $m=0.02$ and $0.03$ and 105 configurations for 
the heavier ones, $m=0.04 - 0.125$. Those bare quark masses 
correspond to mass ratios $m_{\pi}/m_{\rho} 
\approx 0.59 - 0.90$.
All calculations were done on the 600 Gflops QCDSP machine at 
the RIKEN BNL Research Center.

Now let us touch upon technical details. 
We take a linear combination of two quark propagators with periodic 
and anti-periodic boundary condition in the time direction, 
respectively. This procedure enables us to extract the state with 
desired parity even in the baryon spectrum.

\subsection{Parity partner of nucleon: $N^*$}
%
%
\begin{figure}[t]
\label{fig:n*}
\begin{center}
\psfig{figure=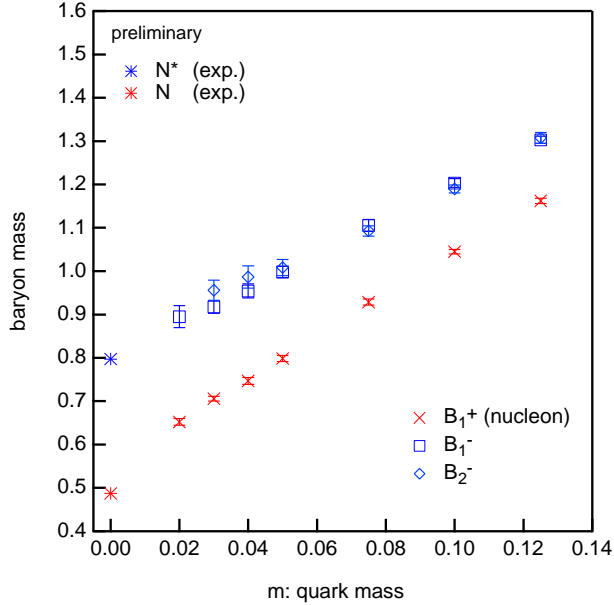,height=8.3cm}
\caption{$N$ ($\times$) 
and $N^*$ ($\Box$ and $\Diamond$) masses versus the quark mass
in lattice units ($a^{-1}\approx 1.9$GeV from $am_{\rho}$=0.400(8) in
the chiral limit). The corresponding experimental values for $N$ and 
$N^*$ are marked with stars. The $N$-$N^*$ mass splitting is clearly observed.}
\end{center}
\end{figure}

\indent\indent
We present mass estimates of the nucleon ($N$) and its parity partner 
($N^*$) obtained by the single mass-fit method applied to the two-point
functions. We first calculate the effective 
masses to find appropriate time ranges for fitting. The effective mass 
plot shows plateaus on the far side of the source 
($t-t_{\rm source}$=13-20) for $N$ and on the near side of the source 
($t-t_{\rm source}$=5-12) for $N^*$.
We choose to fit our baryon propagators from some 
minimum time slice $T_{\rm min}$ to an appropriate maximum time slice 
$T_{\rm max}=20$ for $N$ and $T_{\rm max}\leq 16$ for $N^*$. 
To keep fitting ranges as wide as possible,
$T_{\rm min}$ is varied from $T_{\rm max}-2$ and selected 
under the condition $\chi^2 / N_{\rm DF}\leq 1.0$ where $N_{\rm DF}$
denotes the fitted degree of freedom.
All our fits are reasonable in the sense that 
the confidence-level is larger than 0.3 and estimates 
from the weighted average of the effective mass agree with 
them within errors. 

In Figure 1 
we show the low-lying nucleon spectrum as a function of the quark 
mass, $m$. The nucleon mass is extracted from the $B_{1}^+$ operator.
We omit the point at $m=0.02$ for the operator $B_{2}^-$ since 
a good plateau in the effective mass plot is absent.
$N^*$ mass estimates from $B_{1}^{-}$ and $B_{2}^{-}$ 
operators agree with each other within errors in the whole quark mass 
range, as expected from their common quantum numbers~\cite{Sasaki}.
Note that this result disagrees 
with the $N^*$ spectrum obtained in Ref~16:
we cannot find any discrepancy between mass spectra from $B_{1}^{-}$ 
and $B_{2}^{-}$.
Note also the same signal for $N^*$ is obtained from a mixed 
correlation function 
$\lvac B_{1}^{-}{\bar B}_{2}^{-}+B_{2}^{-}{\bar B}_{1}^{-}\rvac$
in our study. 

The most remarkable feature in Figure 1 is that the $N$-$N^*$ mass splitting 
is observed in the whole range and even for light valence 
quark mass values~\cite{Sasaki}.
This mass splitting grows as the valence quark 
mass decreases~\cite{Sasaki}.
To make this point clear, 
we compare two mass ratios, one from the baryon parity partners
$m_{N^*}/m_N$ and the other from pseudo-scalar and vector mesons 
$m_\pi/m_\rho$ in Figure 2. 
Experimental points are marked with stars, 
corresponding to non-strange (left) and strange (right) sectors. In 
the strange sector we use $\Sigma$ and $\Sigma(1750)$ as baryon 
parity partners and $K$ and $K^*$ for mesons~\cite{PDG}. 
We find the baryon mass ratio clearly grows with decreasing meson mass ratio, 
toward the experimental values~\cite{Sasaki}.

Finally, we evaluate the $N$ and $N^*$ 
mass in the chiral limit.
We simply take a linear extrapolation in 5 lightest quark 
masses for $B_{1}^{+}$ and $B_{1}^{-}$.
We find $m_N$=0.55(1) and $m_{N^*}$=0.80(2)
in lattice units for values in the chiral limit. 
If the scale is set as $a^{-1}\simeq 1.9$ GeV from the $\rho$ mass~\cite{RBC}, 
we obtain $m_{N}\approx1.0$ GeV and $m_{N^*}\approx1.5$ GeV in a
good agreement with the experimental value. 
Above errors do not include systematic errors due to finite volume, 
lattice spacing and quenching effects. Such a systematic study will be 
addressed in future calculations.

%
%
\begin{figure}[t]
\label{fig:ratio}
\begin{center}
\psfig{figure=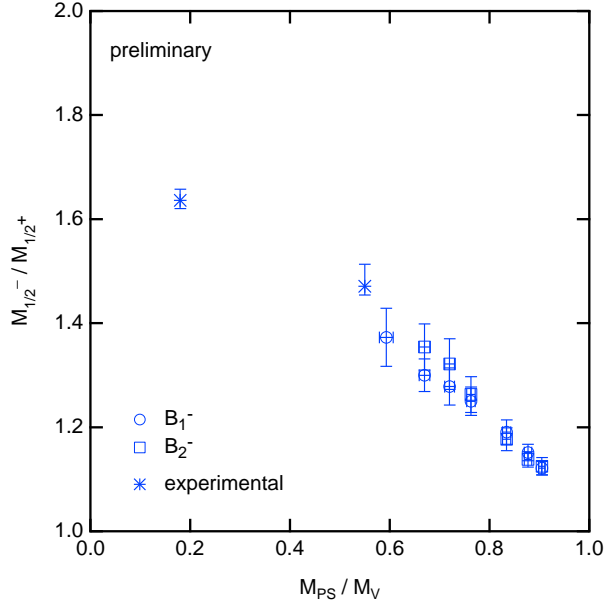,height=8.3cm}
\caption{Mass ratio of 
the negative-parity excited-state and positive-parity ground-state 
baryons versus mass ratio of the pseudoscalar meson and vector meson.
All calculation is done for three degenerate valence quarks.
}
\end{center}
\end{figure}

\subsection{Positive parity excited-state: $N'$}

\indent\indent
In principle, the mass spectrum of the ``first'' excited nucleon can be
derived from the two-mass fits for the nucleon interpolating operator 
$B_{1}^+$. However, large statistics are
required. Also it is difficult to control the systematic errors.
Indeed, several attempts to extract $N'$ mass spectrum failed 
to reproduce experiment. Here, we take another 
approach to $N'$ spectrum. 

As we mentioned, the $B_{2}^+$ operator
vanishes in the non-relativistic limit. Thus we expect that
$B_{2}^+$ weakly couples to the nucleon ground-state since
the non-relativistic description of the nucleon was quite successful
in the naive quark model. On the other hand, nobody has succeeded
in extracting the signal of the nucleon by using the so-called unconventional 
operator $B_{2}^+$ in lattice QCD~\cite{Leinweber}.
This suggests that $B_{2}^+$ has
negligible overlap with the nucleon ground-state and might give us a signal
for an excited-state nucleon.
Of course, this naive expectation is against the common sense that
interpolating operators with the same quantum numbers should give the same mass 
spectrum. However, we find different plateaus in effective mass plots 
from $B_{1}^+$ and $B_{2}^+$ operators~\cite{Sasaki}.

%
%
\begin{figure}[t]
\label{fig:n'}
\begin{center}
\psfig{figure=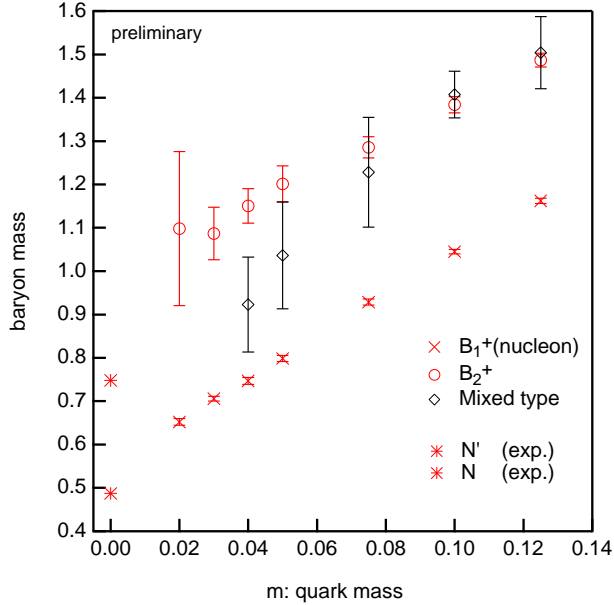,height=8.3cm}
\caption{Mass estimates 
obtained from $B^+_1$ ($\times$), $B^+_2$ ($\circ$) and mixed type 
correlator ($\Diamond$). The experimental values for $N$ and $N'$ 
are marked with stars.}
\end{center}
\end{figure}

We apply the single-mass fit method described above 
to the two-point function composed of $B_{2}^+$.
We find that masses extracted from $B_{1}^{+}$ and $B_{2}^{+}$ 
are quite different as shown in Figure 3.
We conclude that we can identify $B_{2}^{+}$ with the 
``first'' positive-parity excited-state ($N'$) of the nucleon for 
heavy quarks ($m \stackrel{>}{\scriptstyle {\sim}} 0.07$)~\cite{Sasaki}.
The main reason is that we obtain 
mass estimates consistent with
$B_{2}^+$ from the mixed correlation function
$\lvac B_{1}^{+}{\bar B}_{2}^{+}+B_{2}^{+}{\bar B}_{1}^{+}\rvac$.
This suggests $\lvac B_{2}^+ |N\rangle \approx 
0$. Indeed, we see $|\lvac B_{2}^+ |N\rangle / \lvac B_{2}^+ |N' 
\rangle |^2 \leq 10^{-3}$ from the two-mass fits for the $B_{2}^+$ 
correlation at $m=0.10$ and 0.125. This is plausible 
since the operator $B_{2}^{+}$ is expected to couple weakly to the 
ground state of the nucleon as we mentioned earlier~\cite{Leinweber}.
This feature weakens in the lighter quark mass region (from around
$m=0.05$). Unfortunately, we have no data of the mixed type correlation 
for $m=0.02$ and 0.03. To make a definite conclusion about the 
result of $B_{2}^+$, we need further calculations for the light quark mass.

Finally, we want to touch upon the reason why we see a clear $B^+_2$ signal 
for the first time in this study while previous studies failed to do so.
This should be related to mixing induced by explicit chiral symmetry breaking
at finite lattice spacing which is absent in our calculation 
but severe in those calculations.
Although $B_{1}^{+}$ and $B_{2}^{+}$ do not mix in the continuum 
because of different chiral structures\footnote[3]{To speak properly, 
$B_{1}^{+}-B_{2}^{+}$ and $B_{1}^{+}+B_{2}^{+}$ belong to two 
distinct $(\frac{1}{2},0)\oplus(0,\frac{1}{2})$ 
chiral multiplets under $SU(2)_{L}\otimes SU(2)_{R}$~\cite{Cohen}.},
it is known that unwanted mixing between them arises through the 
explicit breaking of chiral symmetry by conventional 
lattice fermions~\cite{Richards}.
On the other hand, this breaking in DWF is expected to be suppressed 
exponentially with $N_{s}$. Thus DWF can drastically
reduce the unwanted mixing~\cite{Aoki}. 
Indeed what we see here suggests such reduction is significant.
As a result, we are able to numerically confirm the expected feature
$\langle 0|B_{2}^{+}|N\rangle\simeq 0$ in the region of heavy 
valence-quark mass. 
\section{Conclusion}

\indent\indent
We studied the mass spectrum of the nucleon excited-states in
quenched lattice QCD using the domain-wall fermions (DWF) 
which preserves the chiral and flavor symmetries almost exactly.  
Most importantly we demonstrated that this method is capable
of calculating the excited-state mass of $N^*$.

We made systematic investigation of the $N^*$ spectrum by using two
distinct interpolating operators, $B_{1}^-$ and $B_{2}^-$.  
We obtained mutually consistent results for the $N^*$ mass spectrum 
from both of them.  This is in contrast with the positive parity 
case as described below.  
In practice $B_{1}^-$ is preferable to $B_{2}^-$ in extracting 
the $N^*$ mass signal.

We found definite mass splitting between
$N^*$ and $N$ in the whole quark mass range studied.  
Furthermore, this splitting grows with decreasing quark mass.  
This is the first time such a remarkable feature has been observed
in any theoretical calculation.
The $N^*$ mass and the $N$-$N^*$ mass splitting in the chiral 
limit obtained by extrapolation is consistent with the experimental 
value within about 5-10\%.  
This is very encouraging for further investigations of $N^*$ physics. 

We observed no signal for the nucleon by using 
the unconventional nucleon operator $B_{2}^+$ which vanishes in the 
non-relativistic limit. Instead, we extracted the mass of 
the first excited nucleon $N'$ for heavy quarks.
We need further study to make a definite conclusion about this subject
for lighter quarks.

\section*{Acknowledgments}

\indent\indent
The author would like to thank Professor V. Burkert and the other
organizers of NSTAR2000 for an invitation.
He is also grateful to Professor L.Ya. Glozman and Professor D.O. Riska
for recommending him to the organizers as an invited speaker in
this workshop.  We thank RIKEN, Brookhaven National Laboratory, and the
U.S.  Department of Energy for providing the facilities essential for the
completion of this work.

\end{document}